\def\apj{{ApJ}}

\def\Msol{\hbox{$\thinspace M_{\odot}$}~}
\def\etal{et al.~\rm}
\def\keV{keV}
\def\Mpc{Mpc}

\documentclass[cup5b]{caps}
\usepackage{graphicx}
\usepackage{amssymb}
\usepackage{ociwsymp3e}  

\HeadText{L. R. Jones et al}  

\def\plottwo#1#2{\centering \leavevmode
\includegraphics[width=.45\columnwidth]{#1} \hfil
\includegraphics[width=.45\columnwidth]{#2}}

\def\plottwo#1#2{\centering \leavevmode
\includegraphics[width=.45\columnwidth]{#1} \hfil
\includegraphics[width=.45\columnwidth]{#2}}

\begin{document}

\pagenumbering{arabic}

\author[]{L. R. JONES$^{1}$, B. J. MAUGHAN$^{1}$, H. EBELING$^{2}$, C. SCHARF$^{3}$,
E. PERLMAN$^{4}$ \and D. LUMB$^{5}$, P. GONDOIN$^{5}$, 
K.O. MASON$^{6}$, F. CORDOVA$^{7}$ and W. C. PRIEDHORSKY$^{8}$
\\
(1) University of Birmingham, UK~
(2) Institute for Astronomy, Hawaii, USA\\
(3) Columbia University, NY, USA~
(4) University of Maryland, MD, USA\\
(5) ESTEC, Noordwijk, The Netherlands~
(6) MSSL, University College London, UK\\
(7) University of California, CA, USA~
(8) Los Alamos National Laboratory, NM, USA }

\chapter{AN XMM and Chandra view of massive clusters of galaxies to z=1}

\begin{abstract}

The X-ray properties of a sample of high redshift (z$>$0.6), massive 
clusters observed with $XMM-Newton$ and $Chandra$  are described,
including two exceptional systems. One, at z=0.89, has an X-ray temperature of
T=11.5$^{+1.1}_{-0.9}$ keV (the highest temperature of any cluster known 
at z$>$0.6), an estimated mass of 
$\approx$(1.4$\pm$0.2)x10$^{15}$ \Msol,
and appears relaxed.  The other,
at z=0.83, has at least three sub-clumps, probably in the process of
merging, and may also show signs of faint filamentary structure at large radii,
observed in X-rays.
In general there is a mix of X-ray morphologies, from those clusters which appear 
relaxed and containing little
substructure to some highly 
non-virialized and probably merging systems. The X-ray gas 
metallicities and gas mass fractions of the relaxed systems are similar
to those of low redshift clusters of the same temperature, suggesting that
the gas was in place, and containing its metals, by z$\approx$0.8.
The evolution of the mass-temperature relation may be consistent with no evolution
or with the ``late formation'' assumption.
The effect of point source contamination in the $ROSAT$ survey from which these 
clusters were selected is estimated, and the 
implications for the $ROSAT$ X-ray luminosity function discussed. 

\end{abstract}

\section{Introduction}

Massive clusters of galaxies are rare objects, forming from the 
high-sigma tail of the cosmological density distribution. Their properties
are powerful probes of cosmology, and give insight into the 
process of structure formation
on large scales. For example, compared to z=0, and
in a $\Lambda$CDM Universe, the number density of virialized
halos of $M>5$x10$^{14}$ \Msol is predicted to be a factor of $\approx$10
lower at z=0.5, or a factor of $\approx$100 lower at z=1, from both simulations 
and the Press-Schechter approximation (eg. Bode \etal 2001). Such a large  
change in number density implies that a significant fraction of the progenitor
systems at high redshift should be unvirialized, perhaps containing
lower mass systems in the process of merging.

The number of genuinely massive clusters known at high redshifts is very
small, and the number of relaxed high redshift massive clusters is even
smaller. Thus the fraction  
which are relaxed, giving direct information
on the epoch  and mode of assembly of massive clusters, is poorly known. 
Relaxed, massive clusters at high redshifts 
also offer the best opportunities for deriving
total masses and gas mass fractions from the X-ray data without 
too many uncertainties due to hydrostatic equilibrium assumptions,
and  also with reasonable signal-to-noise.

Detailed analyses of 3 of the clusters presented here can be found in 
Maughan \etal (2003a,b). We use $\Omega_m$=0.3, $\Omega_{\Lambda}$=0.7 and
H$_0$=70 km s$^{-1}$ unless otherwise stated.

\section{The cluster sample}

The clusters were discovered in a $ROSAT$ serendipitous X-ray survey
(WARPS - Scharf \etal 1997, Jones \etal 1998, Perlman \etal 2002).
Of 16 clusters at z$>$0.6 which form a complete, X-ray selected sample,
10 have been observed with $XMM$ and/or $Chandra$. For 3 of these, the 
$XMM$ observations were of low quality  due to periods of high background.
The remaining observations, however, represent a major step forward 
compared to previous results from $ROSAT$ (see also Rosati \etal, these
proceedings and references therein).

\section{Analysis of the high redshift clusters}

Standard analysis techniques were applied to the $XMM$ and $Chandra$ data,
taking into account the ACIS contamination affecting the low energy efficiency,
and using a local background subtraction method to obtain spectra. 
Images were exposure corrected and adaptively smoothed so that all features are
significant at the 99\% confidence level. 

In general a single temperature was measured 
(but see CLJ1226.9+3332 below for a temperature profile).
Two-dimensional $\beta$-profiles were fit to the images, including PSF blurring, 
to obtain values for 
the core radii and $\beta$.  For the clusters which appeared relaxed, and assuming
spherical symmetry, the gas mass was derived from an integral of the 
density profile given by the $\beta$-profile. The total gravitating mass 
was derived by additionally assuming isothermality and hydrostatic equilibrium.
The total mass within a radius $r$ is then given by

\begin{eqnarray}
\label{egn:hydroeqm}
M(<r) & = & 1.13\times10^{14}\beta\frac{T}{\keV}\frac{r}{\Mpc}\frac{(r/r_c)^2}{1+(r/r_c)^2}M_{\odot}.
\end{eqnarray}

To measure the virial radius, we used eqn (1.1) to find the radius within which the total overdensity
was 200 times the critical density at the redshift of each cluster.
The quoted fluxes, luminosities and masses were measured via an extrapolation from the typical detection radius 
of $\sim$0.3-0.45 $r_{vir}$ to the virial radius. 

\section{Descriptions of the clusters}

\subsection{A massive, relaxed cluster at z=0.89}

\begin{figure}
    \centering
    \includegraphics[width=7cm,angle=0]{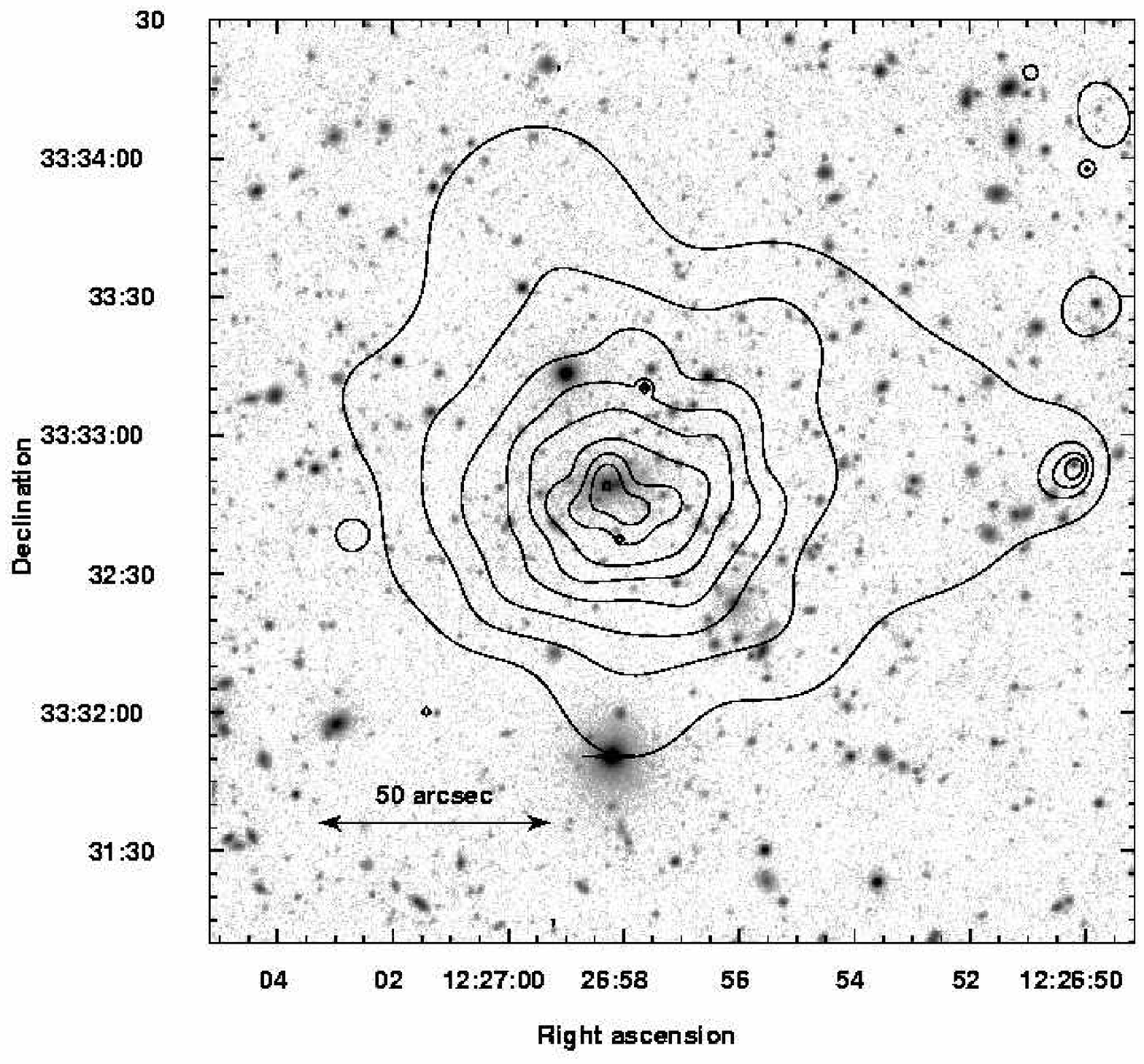}
    \includegraphics[width=6cm,angle=0]{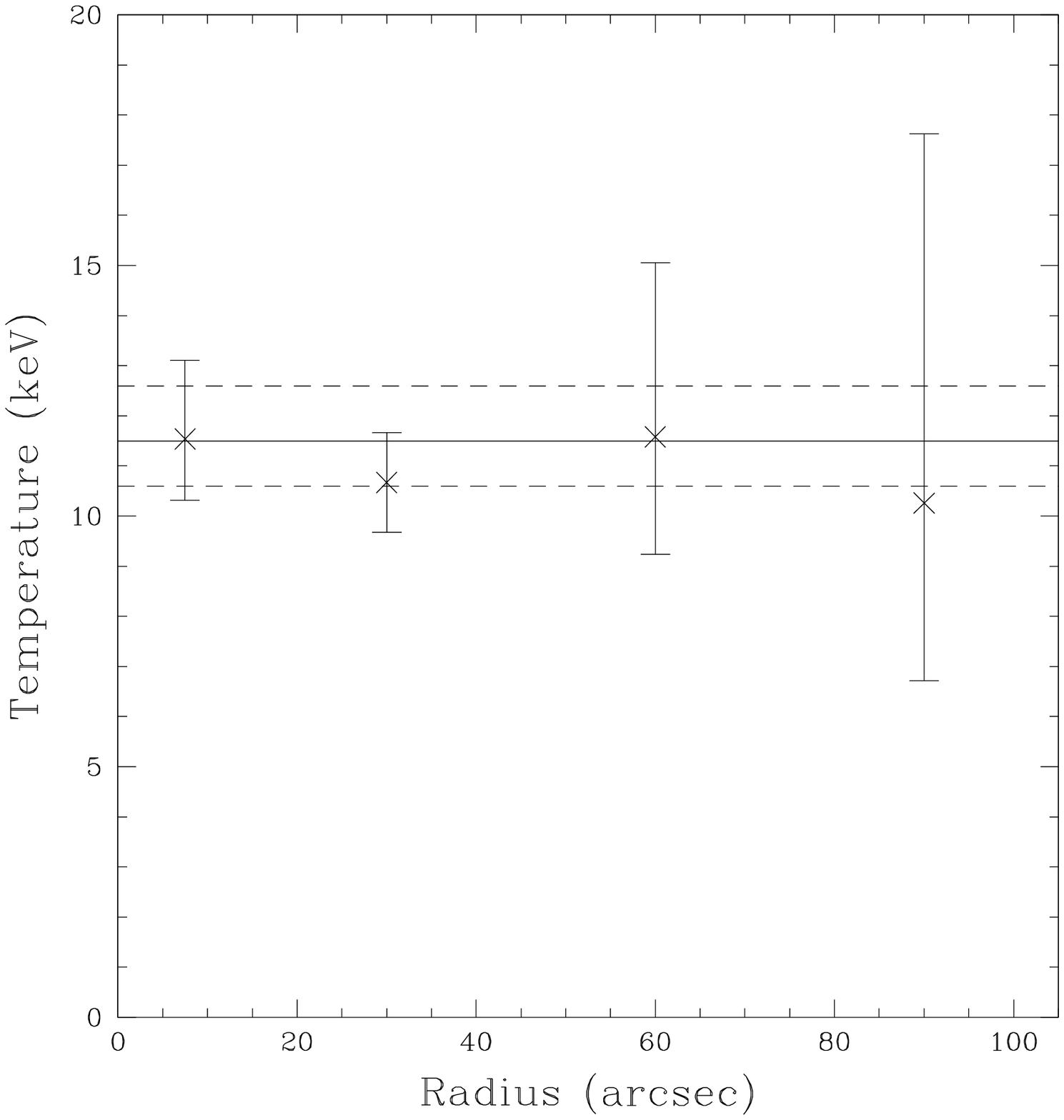}
    \caption{Top - $XMM$ contours (0.3-8 keV) overlayed on a Subaru I-band image of
cluster ClJ1226.9+3332 (z=0.89). The contours are logarithmically spaced.
Bottom - $XMM$ X-ray temperature profile of ClJ1226.9+3332. The solid line is at the best fitting
global temperature, with 1$\sigma$ errors shown by the dashed lines. }
    \label{f12}
 \end{figure}

The discovery of CLJ1226.9+3332 is described by Ebeling \etal (2001).
The $XMM$ image of CLJ1226.9+3332 is shown in Fig \ref{f12}, overlaid on a deep I band image.
The X-ray morphology appears generally
relaxed, in contrast to cluster MS1054-0321, the other high temperature cluster known at z$>$0.8
(Jeltema \etal 2001). The $XMM$ temperature of 11.5$^{+1.1}_{-0.9}$ keV is consistent with the
velocity dispersion of $\approx$1100 km s$^{-1}$, based on 15 galaxy redshifts. The bolometric
X-ray luminosity is 5.4x10$^{45}$ erg s$^{-1}$. We derive a mass of (1.4$\pm$0.2)x10$^{15}$ \Msol
within the virial radius.  Our analysis of a  short $Chandra$ observation confirms the lack of significant
point-source contamination, and the temperature and luminosity, albeit with lower precision (see also
Cagnoni \etal 2001).

A temperature profile (Fig \ref{f12}) is consistent with the cluster being isothermal out to 45\% of
the virial radius. The metal abundance of $Z$=0.33$^{+0.14}_{-0.10} Z_{\odot}$ and gas
mass fraction of 8.6$^{+1.1}_{-1.0}$\% are consistent with those of local clusters of the 
same temperature (see below). 

This cluster is unique in being at high redshift, yet massive and generally relaxed. It must
have been assembled when the age of the Universe was significantly less than 6 Gyr. 
Further details of the $XMM$ analysis and the cosmological implications are given in Maughan \etal 
(2003b).

\subsection{A massive cluster in formation at z=0.83}

\begin{figure}
    \centering
    \includegraphics[width=7cm,angle=0]{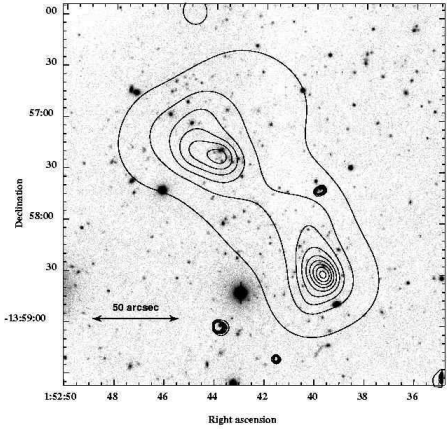}
    \includegraphics[width=7cm,angle=0]{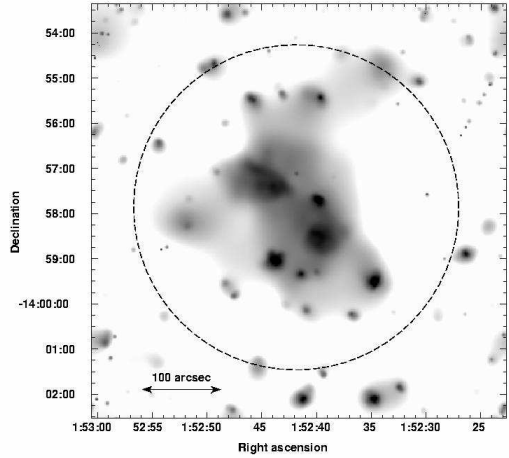}
    \caption{
Top - $Chandra$ contours of ClJ0152.7-1357 (z=0.833).
Bottom - $XMM$ greyscale image of ClJ0152.7-1357. The circle denotes the estimated virial
radius of the final, merged, system. Note the group to the east (at the same redshift as the cluster)
and the filamentary structure, with embedded sources, to the NW.
}
    \label{f34}
 \end{figure}

In stark contrast is the highly unrelaxed, massive cluster ClJ0152.7-1357, described in Ebeling
\etal (2000), Della Ceca \etal (2000) and Maughan \etal (2003a). The $Chandra$ image and a recently
acquired deep $XMM$ image  are shown in Fig \ref{f34}. Based on the $Chandra$ data, we find that 
the cluster consists of two major sub-clumps at very similar redshifts
and of temperatures
5.5$^{+0.9}_{-0.8}$ keV and 5.2$^{+1.1}_{-0.9}$ keV. The total mass of each subcluster is
$\approx$(5-6)x10$^{14}$ \Msol.
A dynamical analysis of the system shows that the subclusters are likely to be gravitationally
bound. When merged, the system mass will be similar to that of the Coma cluster.

The deeper $XMM$ image shows lower surface brightness features. To the east is a low luminosity
system also at the cluster redshift (Demarco \etal 2003, in preparation). To the NW is a possible
filamentary structure with a point source and one or two low luminosity extended sources, probably
groups, embedded within it. This structure falls within a radius of 1.6 Mpc (the dashed circle), the estimated
virial radius of the final merged system, suggesting that it is part of the same formation 
process. The faintest filamentary emission is significant at $>$5$\sigma$ in the soft 0.3-1 keV band
and has a hardness ratio consistent with a temperature of 1$\pm$0.7 keV. We may be observing 
the warm baryons predicted to lie within filaments, based on simulations of structure formation.

\subsection{Other clusters at z$>$0.6}

\begin{figure}
    \centering
\plottwo{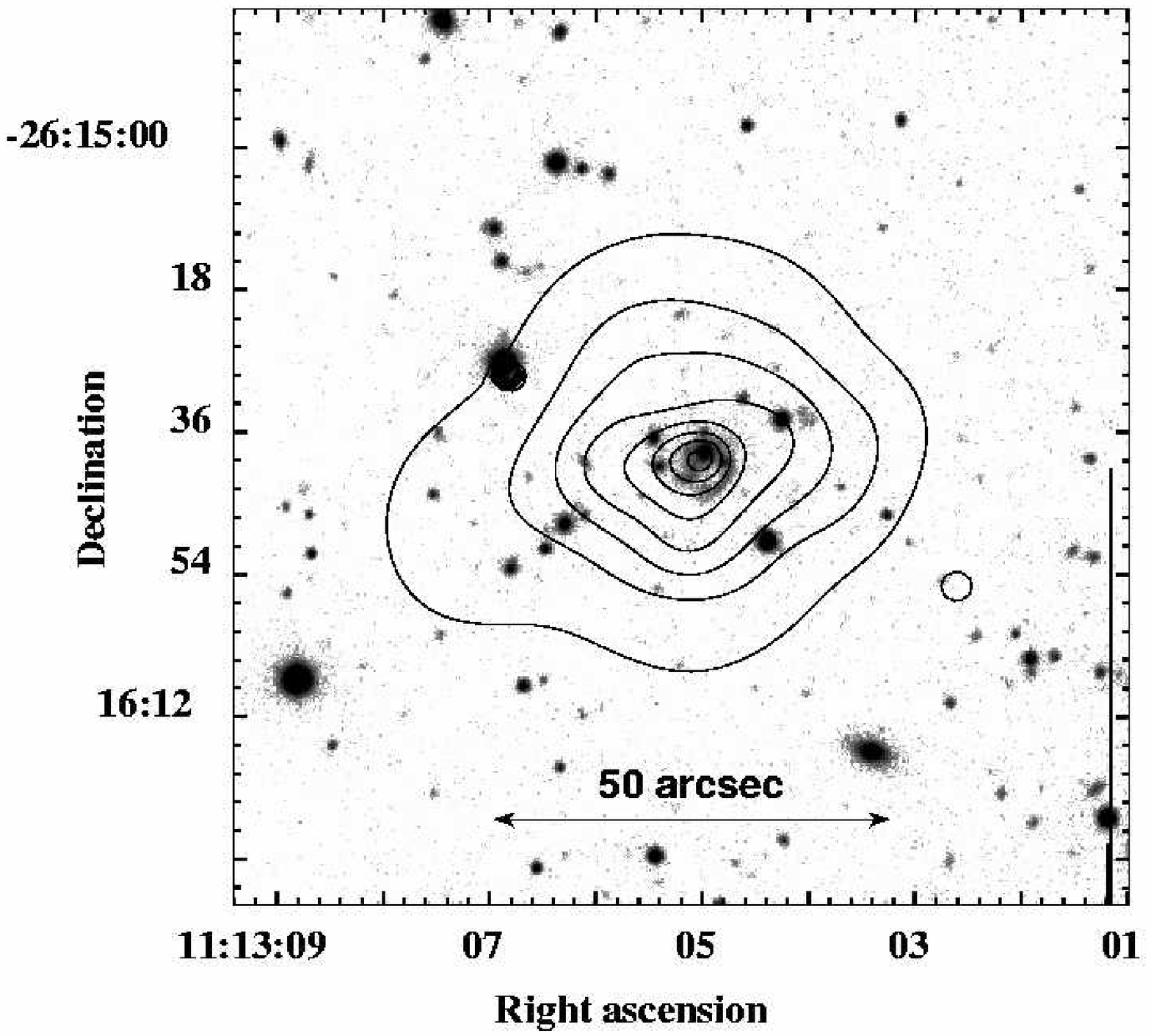}{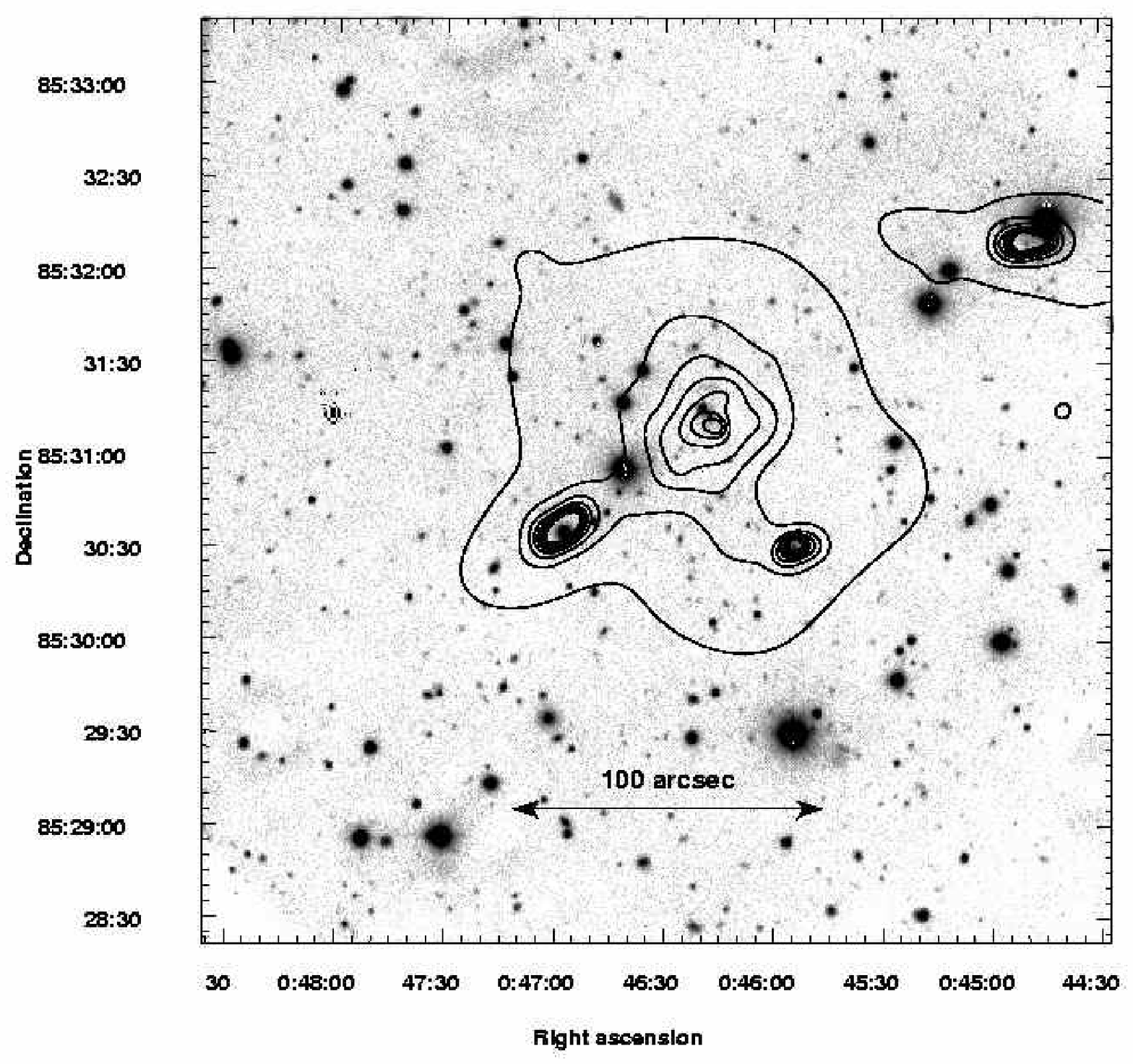}
    \caption{Left - $Chandra$ contours of ClJ1113.1-2615. 
Right - $XMM$ contours of ClJ0046.3+8531  }
    \label{f56}
 \end{figure}

\begin{figure}
    \centering
\plottwo{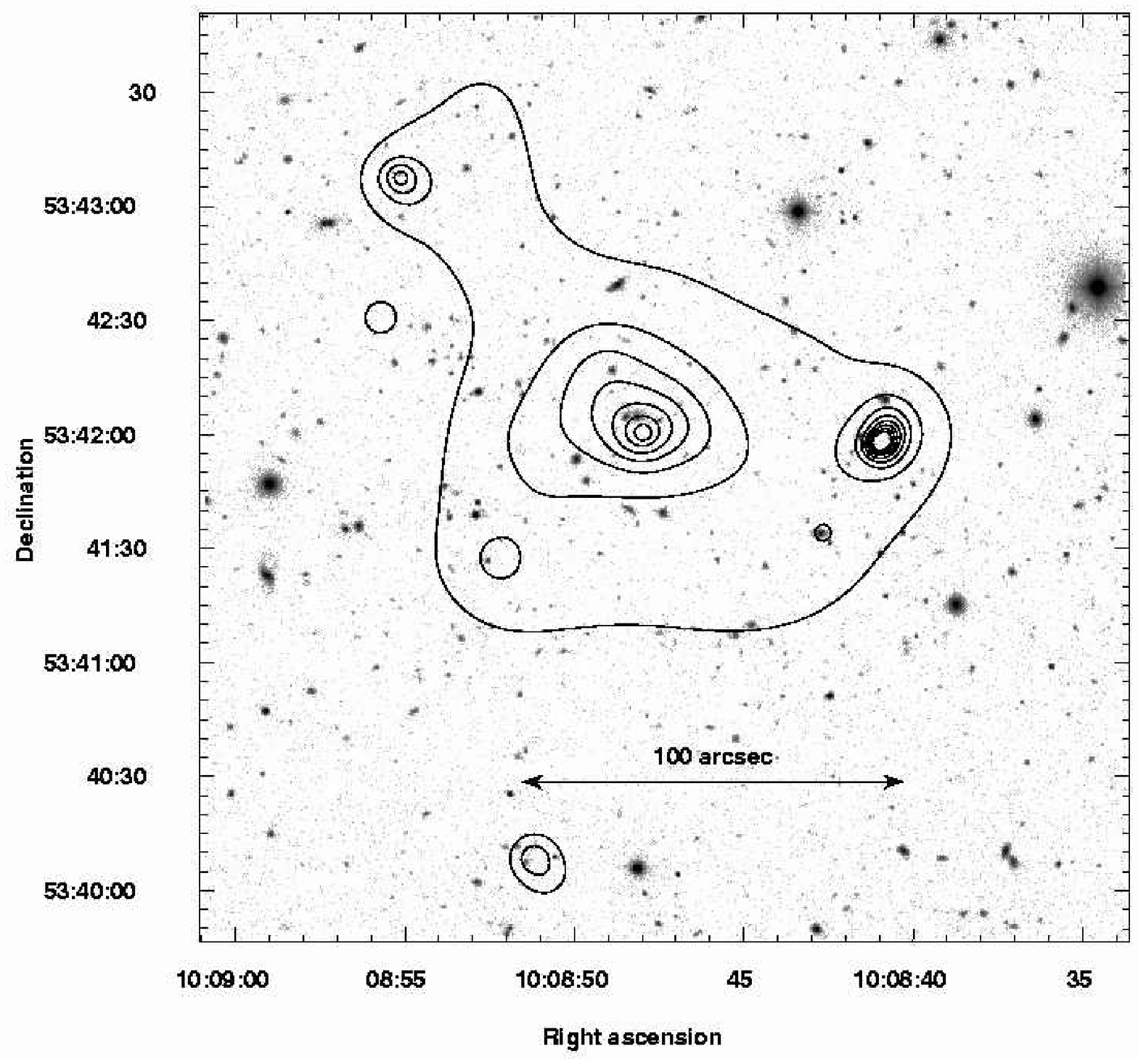}{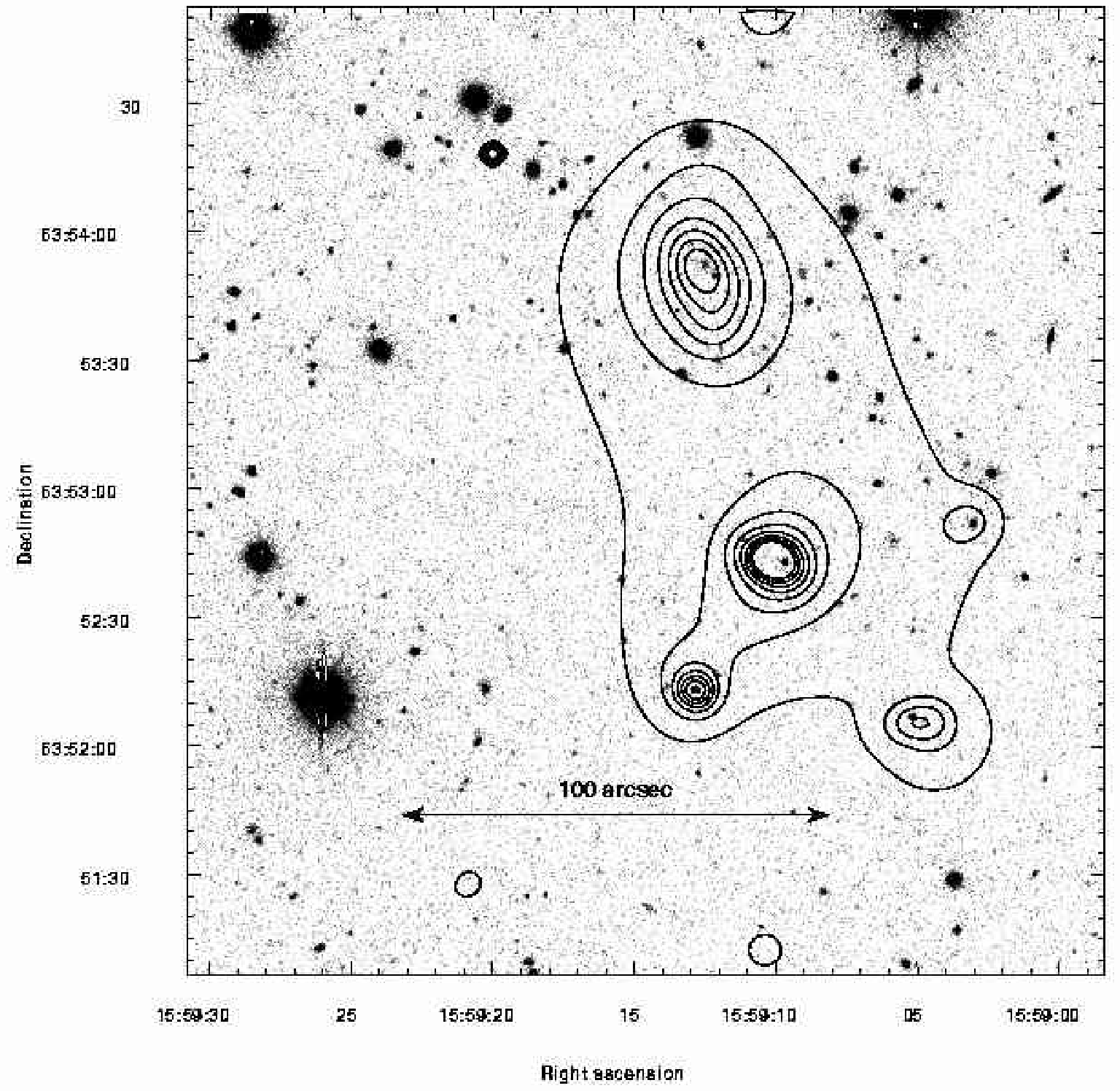}
    \caption{Left - $XMM$ contours of ClJ1008.7+5342 
Right - $XMM$ contours of ClJ1559.1+6353.}
    \label{f78}
 \end{figure}

\begin{figure}
    \centering
   \includegraphics[width=7cm,angle=0]{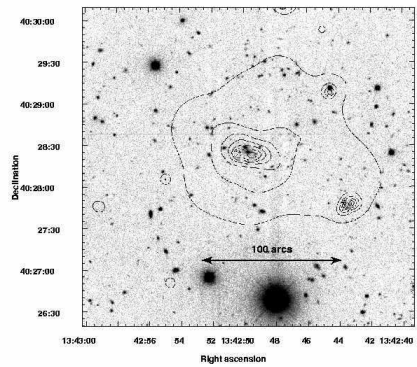}
    \caption{$XMM$ contours of ClJ1342.8+4028.}
    \label{f9}
 \end{figure}

$Chandra$ and $XMM$ contours of the remaining 5 clusters are shown in Figs \ref{f56}, \ref{f78} and 
\ref{f9} in order of
the most relaxed first. There is a wide range of morphologies, and 
some (eg ClJ1559) clearly have bright point X-ray sources nearby. 

\begin{figure}
    \centering
    \includegraphics[width=9cm,angle=270]{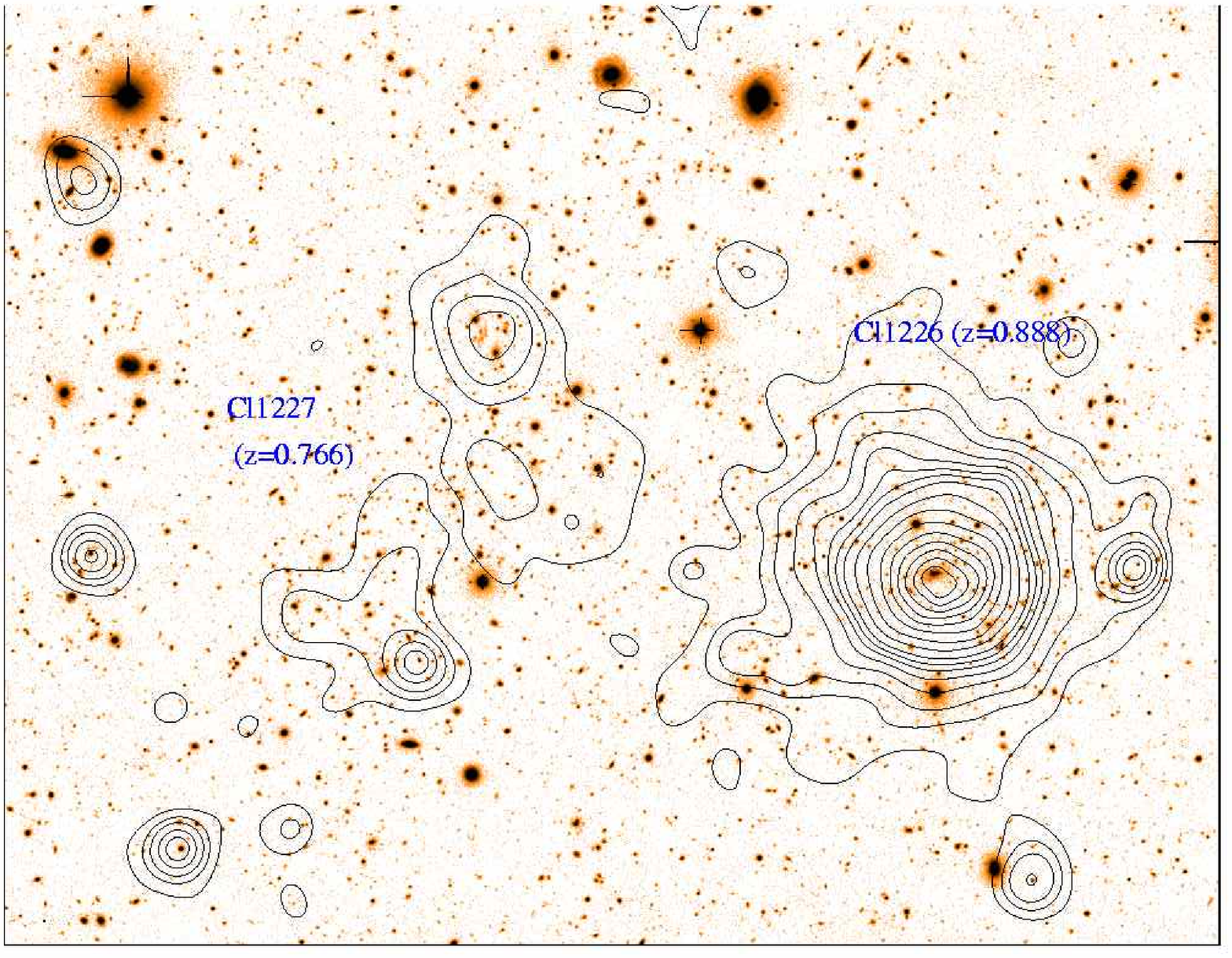}
    \caption{$XMM$ contours of ClJ1227.3+3333 (z=0.766), which is adjacent to 
ClJ1226.9+3332 on the sky, although at a different redshift. The contours are linearly
spaced and the underlying image is in the R band, from Subaru.}
    \label{f10}
 \end{figure}

An additional cluster, ClJ1227.3+3333 (z=0.766) is shown in Fig \ref{f10}. This is an extremely unrelaxed
system. Although not in the original
$ROSAT$ sample (it fell below a window support shadow), it is detected in an $XMM$ image and
is a factor $\sim$12 less luminous than ClJ1226.9+3332.
There is no clear optical or X-ray core, but rather a system of $\approx$4 X-ray subclumps 
separated by $\sim$0.5 Mpc  which
are at a very similar redshift.  There is a red sequence in the galaxy colours. This is probably
a highly non-virialized system seem in formation, yet it is detected in X-rays. 

\section{Metallicities and gas mass fractions}

\begin{figure}
    \centering
    \includegraphics[width=9cm,angle=0]{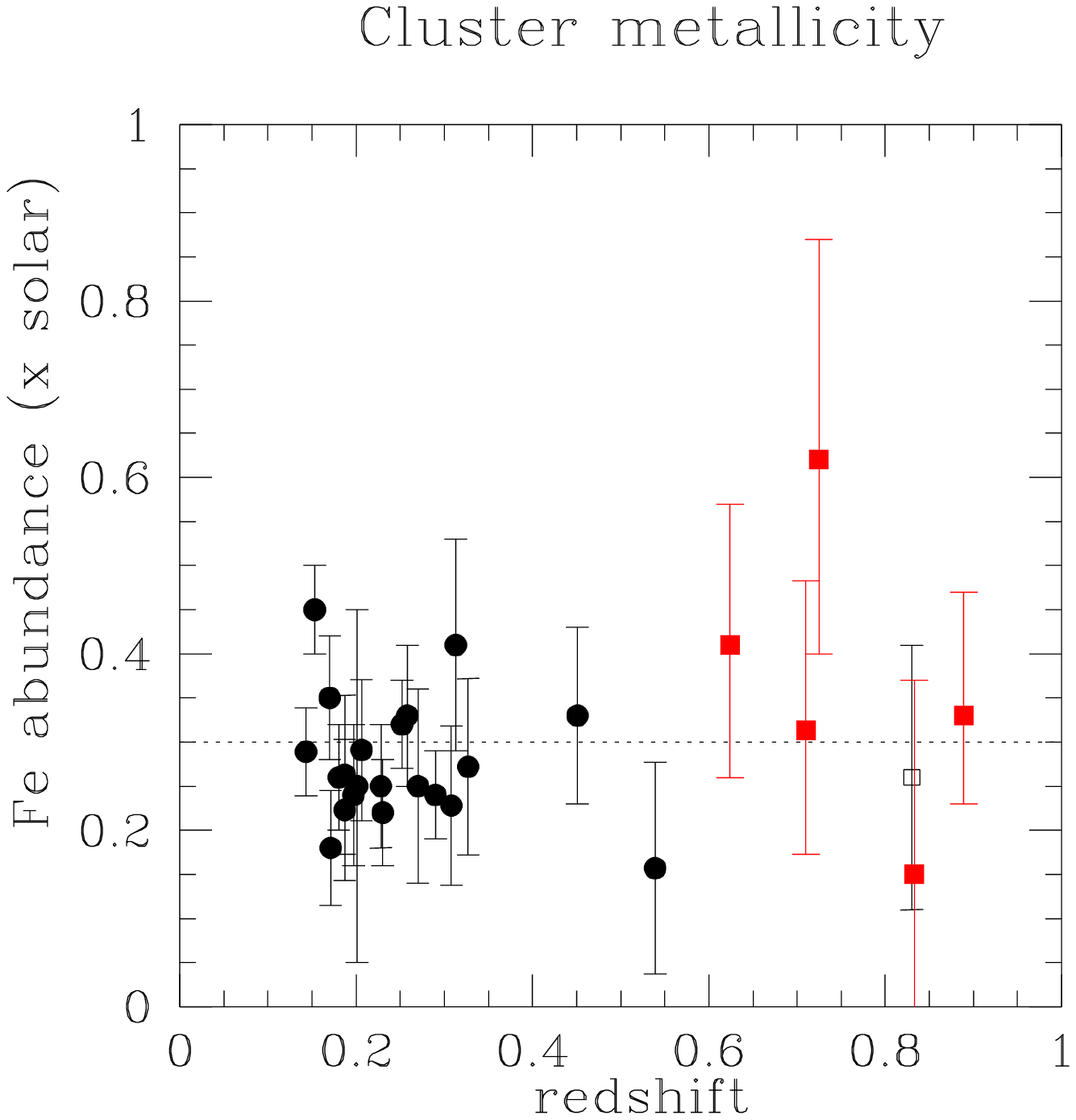}
    \caption{Cluster metallicity vs. redshift. Circles at low redshifts are from Mushotzky \& Loewenstein 
(1997), with 90\% confidence error bars, filled (red) squares represent this work,
with 68\% confidence error bars, and the open square is from Jeltema \etal (2001). }
    \label{f11}
 \end{figure}

Most of the metals in a cluster are in the X-ray gas, and so the evolution of the metallicity of the gas
provides important information on the chemical history of the Universe. In Fig \ref{f11} we show that 
our  measurements, and those of others,  of cluster metallicities up to z=0.9 are consistent with the canonical
value of 0.3 times the solar value found at lower redshifts by Mushotzky \& Loewenstein (1997). 

The gas mass fractions we measure, within a radius of 0.3$r_{vir}$ (where they are the most
reliable) are consistent with those
found by Sanderson \etal (2002) for local clusters of the same temperature. Thus the gas, and the metals 
it contains, were in place by z$\approx$0.8 in at least some massive clusters. 

\section{The evolution of the mass-temperature relation}

\begin{figure}
    \centering
    \includegraphics[width=9cm,angle=0]{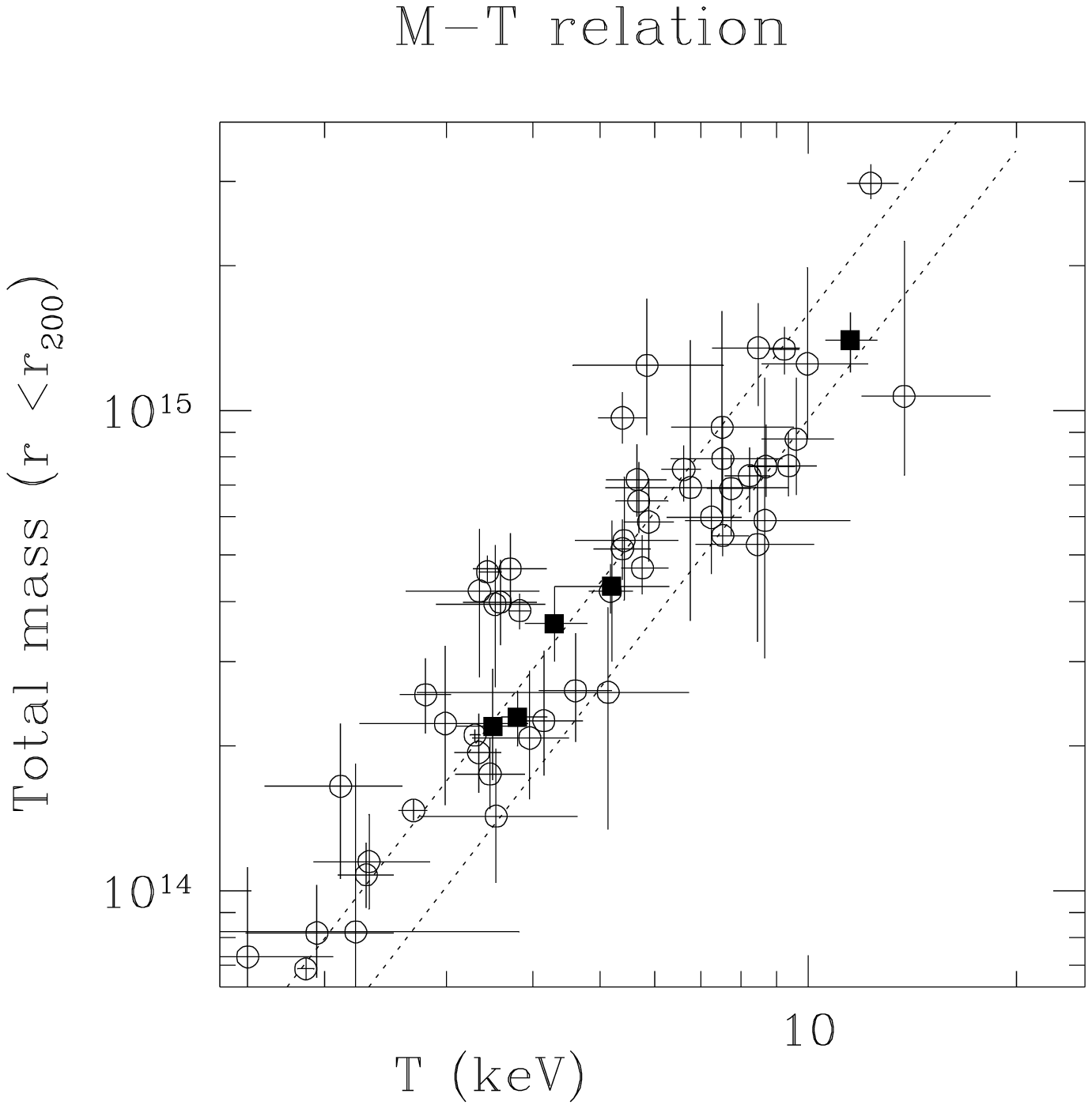}
    \caption{Total mass within $r_{200}$ 
vs. X-ray temperature. Squares (red) represent the relaxed clusters in this work
at high redshifts. Open circles are from Sanderson \etal (2002) at low redshifts. See text for details 
of the measurement methods and the z=0.8 prediction (lower dashed line).
}
    \label{f12a}
 \end{figure}

In Fig \ref{f12a} we show the mass-temperature relation for those high redshift clusters which appear
reasonably relaxed (shown as red squares). The circles are the low redshift data points of 
Sanderson \etal (2002). In both cases the mass is that measured within the virial radius, $r_{200}$, and
in both studies $r_{200}$ is measured from the derived mass profile as the radius within which  the density
is 200 times the critical density at the redshift of observation. The low and high redshift 
samples have different treatments of the temperature profile, however. For the high redshift sample
we (necessarily) assume isothermality. For the low redshift sample, the temperature profile is measured 
and thus a more accurate mass is derived.  

The upper dashed line shows the best fit to the low redshift points. The lower (red) dashed line
shows the expected change in normalisation assuming clusters observed at z=0.8 were formed at z=0.8,
using the relation  $E(z)M\propto T^{3/2}$, where
$E(z)=(1+z)\sqrt{(1+z\Omega_M+\Omega_\Lambda/(1+z)^2-\Omega_\Lambda)}$. It is interesting to note
that all the high redshift clusters except the most massive one are consistent with the low redshift
relation, rather than the z=0.8 prediction. 
It is however the most massive cluster (ClJ1226) which has evidence of isothermality and the most
reliable mass.  The  masses of the other high redshift clusters 
would need to be reduced by $\approx$40\% in order to be consistent with
the z=0.8 prediction.  Sanderson \etal have shown that an incorrect assumption of 
isothermality when clusters are not isothermal produces, on average,
masses which are too high by 30\% at $r_{200}$. This would explain most of the offset we find. 
Further work is in progress, and will be reported in future papers, but
more accurate masses at high redshifts, based on temperature profiles, may be required  before
firm conclusions can be reached.

\section{Point source contamination of $ROSAT$ luminosities}

The $ROSAT$ detections were obtained at $ROSAT$ off-axis angles of up to 15 arcmin, where the 
PSF degrades to $\approx$50 arcsec (FWHM). Thus some point-source contamination is expected 
in all the $ROSAT$ cluster surveys. The degree of contamination  depends on the signal-to-noise ratio
of the detection and the algorithm used for source detection and characterisation.

Based on a preliminary analysis, 2 or 3 out of 9 clusters at z$>$0.6 in the WARPS survey were 
artificially boosted above the survey flux limit by significant point source contamination. Removing these 
sources from the sample, and using the updated $XMM$ and $Chandra$ fluxes, we recomputed the 
X-ray luminosity function at 0.6$<z<$1.1. The resultant shifts to lower space densities were 
within the Poissonian error bars, suggesting that point source contamination can be an important
effect for individual $ROSAT$ measurements, but  the previous  broad conclusions regarding 
cluster evolution are unaffected.


\section{References}

\noindent
Bode, P., Bahcall, N.~A., Ford, E.~B., Ostriker, J.~P., 2001, \apj, 551, 15\\
Cagnoni, I., Elvis, M., Kim, D.-W., Mazzotta, P., Huang, J.-S., Celotti, A., 2001, ApJ, 560, 86\\
Della Ceca, R., Scaramella, R., Gioia, I.M., Rosati, P., Fiore, F., Squires, G., 2000, A\&A, 353, 498\\
Ebeling, H., Jones, L.~R., Perlman, E.~S., Scharf, C.~A., Horner, D.,
Wegner, G., Malkan, M., Mullis, C.R. 2000, ApJ, 534, 133.\\
Ebeling, H., Jones, L.~R., Fairley, B.~W., Perlman, E., Scharf, C., 
Horner, D. 2001, ApJ 548, L23.\\
Jeltema, T.~E., Canizares, C. R., Bautz, M. W., Malm, M. R., Donahue, M., Garmire, G. P., 2001, 
 ApJ, 562, 124\\
Jones, L.R., Scharf, C.A., Ebeling, H., Perlman, E., Wegner, G., Malkan, M., Horner, D.,
1998, ApJ, 495, 100\\
Maughan, B.J., Jones, L.R., Ebeling, H., Perlman, E., Rosati, P., Frye, C., Mullis, C.R.,
2003a, ApJ in press (astro-ph/0301218)\\
Maughan, B.J., Jones, L.R., Ebeling, H., Scharf C., 2003b, in preparation\\
Mushotzky, R. F., Loewenstein, M., 1997, ApJ, 481, L63\\
Perlman, E.S., Horner, D., Jones, L.R., Scharf, C., Ebeling, H., Wegner, G., 
Malkan, M., 2002, ApJS, 140, 265\\
Scharf, C., Jones, L.R., Ebeling, H., Perlman, E., Wegner, G., Malkan, M., 1997, ApJ, 477, 79\\

\section{Acknowledgements} 
We acknowledge useful discussions with Stefano Ettori, Gus Evrard, Monique Arnuad \& Trevor Ponman.

\end{document}